\documentclass[12pt]{article}
\usepackage{epsfig}
\usepackage{slashbox}

\newcommand{\Li}{\mathop{\mathrm{Li}_2}\nolimits}
\newcommand{\agt}{\,\rlap{\lower 3.5 pt \hbox{$\mathchar \sim$}} \raise 1pt
 \hbox {$>$}\,}
\newcommand{\alt}{\,\rlap{\lower 3.5 pt \hbox{$\mathchar \sim$}} \raise 1pt
 \hbox {$<$}\,}

\begin{document}

\title{
\vskip-3cm{\baselineskip14pt
\centerline{\normalsize DESY 07-040\hfill ISSN 0418-9833}
\centerline{\normalsize hep-ph/0703239\hfill}
\centerline{\normalsize March 2007\hfill}}
\vskip1.5cm
Comparative analysis of non-perturbative effects in
$B\to X_ul\overline{\nu}_l$ decays}

\author{Bernd A. Kniehl, Gustav Kramer\\
{\normalsize\it
II. Institut f\"ur Theoretische Physik, Universit\"at Hamburg,
}\\
{\normalsize\it
Luruper Chaussee 149, 22761 Hamburg, Germany.}\\
\\
Ji-Feng Yang\\
{\normalsize\it
Department of Physics, East China Normal University,
}\\
{\normalsize\it
3663th Zhong Shan Bei Road, Shanghai 200062, China}}

\date{}

\maketitle

\begin{abstract}
In order to extract the Cabibbo-Kobayashi-Maskawa matrix element $|V_{ub}|$
from $B\to X_ul\overline{\nu}_l$ decays, the overwhelming background from
$B\to X_cl\overline{\nu}_l$ decays must be reduced by appropriate acceptance
cuts.
We study the non-perturbative effects due to the motion of the $b$ quark inside
the $B$ meson on the phenomenologically relevant decay distributions of
$B\to X_ul\overline{\nu}_l$ in the presence of such cuts in a comparative
analysis based on shape functions and the parton model in the light-cone
limit.
Comparisons with recent data from the CLEO, BABAR, and BELLE collaborations
favor the shape-function approach.

\medskip

\end{abstract}

\newpage

\section{\label{sec:one}Introduction}

To test the predictions of the Standard Model for the simultaneous violation
of the charge conjugation and parity (CP) symmetries in $B$-meson decays, it
is very important to know the matrix element $|V_{ub}|$ of the
Cabibbo-Kobayashi-Maskawa quark-mixing matrix \cite{Cabibbo:1963yz} very
accurately.
The uncertainties in existing measurements, by the CLEO
\cite{Bornheim:2002du}, BABAR
\cite{Aubert:2003zw,Aubert:2005im,Aubert:2005mg,Aubert:2006qi}, and BELLE
\cite{Kakuno:2003fk,Limosani:2005pi,Bizjak:2005hn} collaborations,
are dominantly due to uncertainties in the theoretical calculation of partial
decay rates to be compared with the experimental measurements.
Experimentally, the inclusive rate $\Delta\Gamma_{ul\nu}(\Delta\Phi)$ of
$B\to X_ul\overline{\nu}_l$ decays in a restricted region $\Delta\Phi$ of
phase space is measured, where the dominant charm background is suppressed and
theoretical uncertainties are reduced.
The theoretical factor $R(\Delta\Phi)$ directly relates the inclusive rate to
$|V_{ub}|$ without extrapolation to the full phase space, as
\begin{equation}
|V_{ub}|^2=\frac{\Delta\Gamma_{ul\nu}(\Delta\Phi)}{R(\Delta\Phi)}.
\end{equation}
The uncertainties in the calculation of $R(\Delta\Phi)$ dominantly originate
from the modeling of the Fermi motion of the $b$ quark inside the $B$ meson.
Most of the recent analyses towards the determination of $|V_{ub}|$ from
measurements of $\Delta\Gamma_{ul\nu}(\Delta\Phi)$
\cite{Aubert:2005im,Aubert:2005mg,Limosani:2005pi,Bizjak:2005hn} rely on the
calculation of $R(\Delta\Phi)$ by Lange {\it et al.}\ \cite{Lange:2005yw}.
They use the so-called shape-function (SF) scheme, which is an extended
version of the original SF approach \cite{Neubert:1993ch,Bigi:1993ex} with
many effects due to renormalization-group-improved perturbation theory,
higher-order power corrections from subleading SF terms, {\it etc.}
But there are many more approaches known for describing the non-perturbative
$B\to b$ transition.
We mention the Altarelli-Cabibbo-Corbo-Maiani-Martinelli (ACCMM) model
\cite{Altarelli:1982kh}, one of the oldest models to describe the motion of
the $b$ quark inside the $B$ meson.
In this model, it is assumed that the $B$ meson consists of the $b$ quark and
a spectator quark, with definite mass $m_{\rm spec}$ and momentum
$p_{\rm spec}$, which is considered quasi-free.
The $b$ quark is treated as a virtual particle with a mass depending on
$p_{\rm spec}$.
Another popular model for describing the non-perturbative $B\to b$ transition
is the model of Bareiss, Jin, Palmer, and Paschos based on the parton model
approach in the light-cone (LC) limit \cite{Bareiss:1989my,Jin:1995hi}.
All these models, including the SF models, contain phenomenological functions
of the respective variables describing the motion of the $b$ quark inside the
$B$ meson with parameters fitted to the $b$-quark mass and one or two
characteristic moments of these functions.
Another approach tries to avoid these non-perturbative functions by assuming
that the fragmentation of the $B$ meson into the $b$ quark and the spectator
quark can be described as a radiation process off the $b$ quark with a proper
coupling inserted in the standard soft-gluon resummation formula
\cite{Aglietti:2006yb}.
For a similar approach, referred to as dressed-gluon exponentiation (DGE)
in the literature, see Ref.~\cite{Andersen:2005mj}.

Given the variety of approaches for treating the non-perturbative transition,
it is desirable to make an attempt to compare these approaches with respect to
their predictions for $R(\Delta\Phi)$ and other physical observables.
In this work, we shall make such a comparison between the simple SF approach
and the parton model approach in the LC limit, which we shall refer to as the
LC approach in the following.
Such a comparison of the parton model and the ACCMM model has already been
done some time ago in Ref.~\cite{Kim:1998wx}.

The outline of this work is as follows.
In Sect.~\ref{sec:two}, we give a short introduction to the SF and LC
approaches.
Section~\ref{sec:three} contains the results for $R(\Delta\Phi)$ for three
choices of $\Delta\Phi$ underlying recent experimental measurements by BABAR
and BELLE.
In addition to $R(\Delta\Phi)$, we also present in Sect.~\ref{sec:three}
distributions in several kinematical variables and compare them with measured
differential decay distributions.
Section~\ref{sec:four} contains a summary and the conclusions.

\section{\label{sec:two}Theoretical ingredients}

\subsection{\label{sec:twoa}Perturbative differential decay rate}

The differential decay width of $B\to X_ul\overline{\nu}_l$ has been
calculated up to first order in the strong-coupling constant $\alpha_s$ by De
Fazio and Neubert \cite{DeFazio:1999sv} using a fictitious gluon mass to
regulate soft and collinear gluon contributions.
This result has been confirmed using dimensional regularization for the soft
and collinear singularities in Ref.~\cite{kim} and by us.
The quantity of interest is the triple differential decay rate
$d^3\Gamma/(dx\,dz\,d\hat p^2)$ of
\begin{equation}
b(p_b)\to X_u(p)+l(p_l)+\overline{\nu}_l(p_\nu),
\end{equation}
where $X_u=u$ or $X_u=u+g$ in the case of single-gluon emission and the
assigned four-momenta are displayed in parentheses.
Introducing $p=p_u+p_g$ and $q=p_l+p_\nu$, we have $p_b=p+q$.
The variables $x$, $z$, and $\hat p^2$ are defined as
\begin{equation}
x=\frac{2p_b\cdot p_l}{m_b^2},\qquad
z=\frac{2p_b\cdot p}{m_b^2},\qquad
\hat p^2=\frac{p^2}{m_b^2},
\end{equation}
and take the values
\begin{equation}
0\le x\le 1,\qquad
\overline{x}\le z\le1+\overline{x},\qquad
\max(0,z-1)\le\hat p^2\le\overline{x}(z-\overline{x}),
\end{equation}
where $\overline{x}=1-x$.
The variable $\hat p^2$ measures the invariant mass square of the hadronic
system $X_u$ in units of $m_b^2$, while, in the $b$-quark rest frame, $x$ and
$z$ correspond to the energies of $l$ and $X_u$ in units of $m_b/2$,
respectively.
For fixed values of $z$ and $\hat p^2$, $\overline{x}$ varies in the range
\begin{equation}
\frac{z-\sqrt{z^2-4\hat p^2}}{2}\le\overline{x}
\le\frac{z+\sqrt{z^2-4\hat p^2}}{2}.
\end{equation}
Doubly and singly differential decay distributions are obtained by
appropriately integrating over $d^3\Gamma/(dx\,dz\,d\hat p^2)$.
The simplest distribution is the spectrum in $x$, which reads
\cite{DeFazio:1999sv,Jezabek:1988ja}:
\begin{equation}
\frac{1}{\Gamma_0}\,\frac{d\Gamma}{dx}=2x^2(3-2x)\left[1-
\frac{C_F\alpha_s}{2\pi}G(x)\right],
\label{eq:dgdx}
\end{equation}
where $C_F=4/3$,
\begin{equation}
\Gamma_0=\frac{G_F^2|V_{ub}|^2m_b^5}{192\pi^3},
\label{eq:gam0}
\end{equation}
with $G_F$ being Fermi's constant, is the total decay rate at leading order
(LO) and
\begin{eqnarray}
G(x)&=&\ln^2(1-x)+2\Li(x)+\frac{2}{3}\pi^2+\frac{82-153x+86x^2}{12x(3-2x)}
\nonumber\\
&&{}+\frac{41-36x+42x^2-16x^3}{6x^2(3-2x)}\ln(1-x),
\end{eqnarray}
with $\Li$ being the Spence function.
By integrating over $x$, one obtains the well-known ${\cal O}(\alpha_s)$
formula for the total decay rate of $b\to X_ul\overline{\nu}_l$:
\begin{equation}
\Gamma=\Gamma_0\left[1-\frac{C_F\alpha_s}{2\pi}
\left(\pi^2-\frac{25}{4}\right)\right].
\label{eq:gamtot}
\end{equation}
Formulas for other doubly differential distributions like
$d^2\Gamma/(dz\,d\hat p^2)$ and $d^2\Gamma/(dx\,dz)$ or singly differential
distributions like $d\Gamma/dz$ and $d\Gamma/d\hat p^2$ may be found in
Ref.~\cite{DeFazio:1999sv}.
From $d^2\Gamma/(dz\,d\hat p^2)$, also the distribution in the hadronic
invariant mass $M_X$ can be calculated.
In the heavy-quark limit, where $p_B=(M_B/m_b)p_b$, one has
\begin{equation}
M_X^2=\hat p^2m_b^2+zm_b\overline{\Lambda}+\overline{\Lambda}^2,
\end{equation}
where $\overline{\Lambda}=M_B-m_b$.

\subsection{\label{sec:twob}SF approach}

In kinematic regions close to the phase space boundaries, the perturbative
spectra are infrared sensitive and expected to receive large non-perturbative
corrections.
Such corrections are due to the motion of the $b$ quark inside the $B$ meson
and are usually referred to as Fermi-motion corrections
\cite{Altarelli:1982kh}.
In the singly differential spectra, such regions are
$1-x={\cal O}(\Lambda_{\rm QCD}/m_b)$ for the charged-lepton energy spectrum,
$1-z={\cal O}(\Lambda_{\rm QCD}/m_b)$ for the hadronic energy spectrum, and
the low-hadronic-mass region $M_X^2={\cal O}(\Lambda_{\rm QCD}m_b)$, where
$\Lambda_{\rm QCD}\approx0.5$~GeV is the asymptotic scale parameter of QCD.

One popular method to incorporate Fermi-motion effects is the introduction of
a SF $F(k_+)$, which is supposed to describe light-cone momentum distribution
of the $b$ quark inside the $B$ meson \cite{Neubert:1993ch,Bigi:1993ex}.
The component $k_+$ of the $b$-quark light-cone momentum varies between $-m_b$
and $\overline{\Lambda}$ with a distribution centered around $k_+=0$ and having
a characteristic width of ${\cal O}\left(\overline{\Lambda}\right)$.
The physical $B$-meson decay distributions are calculated from a convolution
of the perturbative $b$-quark decay spectra with $F(k_+)$.
This is done by replacing the $b$-quark mass by the momentum-dependent mass
$m_b+k_+$.
Similarly, the parameter $\overline{\Lambda}$ is replaced by
$\overline{\Lambda}-k_+$ \cite{Neubert:1993ch}.
Introducing $q_+=\overline{\Lambda}-k_+$, the charged-lepton energy
distribution, for example, is modified to become \cite{DeFazio:1999sv}
\begin{equation}
\frac{d\Gamma}{dE_l}(B\to X_ul\overline{\nu}_l)=2\int_0^{M_B-2E_l}dq_+
\frac{F\left(\overline{\Lambda}-q_+\right)}{M_B-q_+}\,\frac{d\Gamma(x_q)}{dx},
\label{eq:dgde}
\end{equation}
where $d\Gamma/dx$ is the perturbative spectrum given in Eq.~(\ref{eq:dgdx}),
$x_q=2E_l/(M_B-q_+)$, and the charged-lepton energy $E_l$ varies in the range
$0\le E_l\le M_B/2$.
The analogous formulas for the distributions in the total hadronic energy and
the hadronic mass may be found in Ref.~\cite{DeFazio:1999sv} and will not be
repeated here.
Since we wish to calculate the fractional decay rate with cuts on $E_l$ and
$M_X$, we need the doubly differential distribution $d^2\Gamma/(dE_l\,dM_X)$.
This and the triply differential distribution $d^2\Gamma/(dE_l\,dM_X\,dq^2)$
are derived analogously to Eq.~(\ref{eq:dgde}).
After the implementation of the SF, the kinematic variables take values in the
entire phase space determined by hadron kinematics.
For example, the maximum lepton energy is $E_l^{\rm max}=M_B/2$, whereas it is
equal to $m_b/2$ for the phase space of the perturbative decay rate.

Several functional forms of $F(k_+)$ are available in the literature.
They are constrained through moments $A_n=\langle k_+^n\rangle$ of $F(k_+)$,
which are related to the forward matrix elements of local operators on the
light cone \cite{Lange:2005yw}.
The first three moments are
\begin{equation}
A_0=1,\qquad A_1=0,\qquad A_2=\frac{\mu_\pi^2}{3},
\end{equation}
where $\mu_\pi^2$ is the average momentum square of the $b$ quark inside the
$B$ meson \cite{Falk:1992wt}.
In our analysis, we adopt the exponential form \cite{Kagan:1998ym}
\begin{equation}
F(k_+)=N\overline{\Lambda}^{-c}(\overline{\Lambda}-k_+)^c
{\rm e}^{(1+c)k_+/\overline{\Lambda}},
\label{eq:shape}
\end{equation}
which obeys $A_1=0$ if one neglects terms exponentially small in
$m_b/\overline{\Lambda}$.
The condition $A_0=1$ fixes the normalization factor $N$, and the parameter
$c$ is related to the second moment as
\begin{equation}
A_2=\frac{\overline{\Lambda}^2}{1+c}.
\label{eq:a2}
\end{equation}
So, the $b$-quark mass $m_b$ (or $\overline{\Lambda}$) and the parameter $c$
(or $\mu_\pi^2$) are the two input parameters of $F(k_+)$.
Our choice of $\overline{\Lambda}$ and $\mu_\pi^2$ will be specified in
Sect.~\ref{sec:three}, when we present our results for the cut-dependent
partial decay rates $R(\Delta\Phi)$.

\subsection{\label{sec:twoc}LC approach}

Since the $B$ meson is heavy, the momentum transferred in the decay to the
final state is, in most regions of phase space, much larger than the energy of
hadronic binding, which is of ${\cal O}(\Lambda_{\rm QCD})$.
This suggests that the semileptonic decay of the $B$ meson can be treated in a
way analogous to deep-inelastic scattering (DIS) in lepton-proton collisions.
There, LC dynamics dominates DIS and leads to the well-known scaling of the
DIS structure functions.
This is implemented in the parton model, where in LO the structure functions
are given by the parton distribution functions.
These are functions of the scaling variable $\xi$, which relates the parton
four-momentum $p_q=\xi p_p$ to the proton four-momentum $p_p$.
In an analogous manner, the hadron decay process $B\to X_ul\overline{\nu}_l$
is modeled by convoluting the parton decay process
$b\to X_ul\overline{\nu}_l$ with the distribution function $f(\xi)$ of the
momentum $p_b=\xi p_B$ of the $b$-quark inside the $B$ meson according to
\begin{equation}
d\Gamma(B\to X_ul\overline{\nu}_l)=\int d\xi\,f(\xi)
\left.d\Gamma(b\to X_ul\overline{\nu}_l)\right|_{p_b=\xi p_B}.
\end{equation}
This has the consequence that $m_b=\xi M_B$ is also smeared with the variable
$\xi$.
The distribution function $f(\xi)$ can be expressed in terms of the matrix
element of the LC bilocal $b$-quark operator between $B$-meson states as
\cite{Jin:1999rs}
\begin{equation}
f(\xi)=\frac{1}{4\pi M_B^2}\int d(y\cdot p_B)\,{\rm e}^{i\xi y\cdot p_B}
\langle B|\overline{b}(0)\gamma\cdot p_B(1-\gamma_5)U(0,y)b(y)
|B\rangle|_{y^2=0},
\label{eq:fxi}
\end{equation}
where $U(0,y)$ is a gauge link associated with the background gluon field that
ensures the gauge invariance of $f(\xi)$.
The distribution function $f(\xi)$ is positive and has non-zero values for
$0\le\xi\le1$ only.
It fulfills three sum rules \cite{Jin:1999rs}.
One of them is due to $b$-quark number conservation and reads
\begin{equation}
\int_0^1d\xi\,f(\xi)=1.
\label{eq:nor}
\end{equation}
Reducing the bilocal operator in Eq.~(\ref{eq:fxi}) to a local one with the
help of the operator product expansion \cite{Jin:1999rs} in heavy-quark
effective theory (HEQT), one obtains two more sum rules.
They determine, up to ${\cal O}(\Lambda_{\rm QCD}^2/m_b^2)$, the mean value
$\mu$ and the variance $\sigma^2$ of $f(\xi)$, which characterize the position
of the maximum and the width of the distribution \cite{Jin:1999rs}:
\begin{eqnarray}
\mu&=&\int_0^1d\xi\,\xi f(\xi)=\frac{m_b}{M_B}\left(1+\frac{5}{3}E_b\right),
\nonumber\\
\sigma^2&=&\int_0^1d\xi\,(\xi-\mu)^2f(\xi)
=\frac{m_b^2}{M_B^2}\left[\frac{2}{3}K_b-\left(\frac{5}{3}E_b\right)^2\right],
\label{eq:musi}
\end{eqnarray}
where
\begin{eqnarray}
G_b&=&-\frac{1}{2M_B}\langle B|\overline{h}
\frac{g_sG_{\alpha\beta}\sigma^{\alpha\beta}}{4m_b^2}h|B\rangle,
\nonumber\\
K_b&=&-\frac{1}{2M_B}\langle B|\overline{h}
\frac{(iD)^2}{2m_b^2}h|B\rangle,
\nonumber\\
E_b&=&G_b+K_b.
\end{eqnarray}
Here, $g_s=\sqrt{4\pi\alpha_s}$, $h$ is the $b$-quark field, $G_{\alpha\beta}$
is the field strength tensor of the strong force, and $D$ is the covariant
derivative involving the gluon field.
The matrix elements $G_b$ and $K_b$ measure the chromomagnetic energy due to
the $b$-quark spin and the kinetic energy of the $b$ quark inside the $B$
meson, respectively.
Both are dimensionless HQET parameters of
${\cal O}(\Lambda_{\rm QCD}^2/m_b^2)$ and are often related to the alternative
parameters
\begin{eqnarray}
\lambda_1&=&-2m_b^2K_b,
\nonumber\\
\lambda_2&=&-\frac{2}{3}m_b^2G_b.
\end{eqnarray}
The parameter $\lambda_2$ can be extracted from the $B^*$--$B$ mass splitting
yielding\break
$\lambda_2=\left(M_{B^*}^2-M_B^2\right)/4\approx0.12$~GeV$^2$.
Values for $\lambda_1=-\mu_\pi^2$, introduced earlier, will be specified in
Sect.~\ref{sec:three}, when we present our results.
If we introduce these two parameters in Eq.~(\ref{eq:musi}), we have
\begin{eqnarray}
\mu&=&\frac{m_b}{M_B}\left(1-\frac{5(\lambda_1+3\lambda_2)}{6m_b^2}\right),
\nonumber\\
\sigma^2&=&
\frac{m_b^2}{M_B^2}\left[-\frac{\lambda_1}{3m_b^2}
-\left(\frac{5(\lambda_1+3\lambda_2)}{6m_b^2}\right)^2\right].
\label{eq:musig}
\end{eqnarray}
Of course, the three parameters $m_b$, $\lambda_1$, and $\lambda_2$ only
constrain the position of the maximum and the width of the distribution.
For numerical evaluations, one needs the whole function $f(\xi)$, for which we
adopt the ansatz \cite{Jin:1995hi,Jin:1997aj}
\begin{equation}
f(\xi)=N\frac{\xi(1-\xi)}{a^2+(\xi-b)^2}\theta(\xi)\theta(1-\xi).
\label{eq:ans}
\end{equation}
The parameters $a$ and $b$ are determined from the values of $\mu$ and
$\sigma^2$.
The normalization factor $N$ is fixed by Eq.~(\ref{eq:nor}).
For $b=m_b/M_B$ and $a\to0$, Eq.~(\ref{eq:ans}) becomes a delta function,
namely $f(\xi)=\delta(\xi-m_b/M_B)$.
In the following, we shall always use $\lambda_1$ and $\lambda_2$ as input to
determine $a$ and $b$ via Eq.~(\ref{eq:musig}).

\section{\label{sec:three}Numerical results}

The large background from $B\to X_cl\overline{\nu}_l$ is the main limitation
for measuring $|V_{ub}|$.
To reject this background, kinematic cuts have to be applied.
Depending on these cuts, the acceptance for $B\to X_ul\overline{\nu}_l$ decays
is reduced.
With such acceptance cuts applied, the calculation of the
$B\to X_ul\overline{\nu}_l$ decay rate is more complicated and, in particular,
influenced much more strongly by the modeling of the non-perturbative
$B\to b$ transition than without cuts.

In recent experimental analyses, four types of cuts have been introduced to
separate $B\to X_ul\overline{\nu}_l$ decays from the much more abundant
$B\to X_cl\overline{\nu}_l$ decays.
First, various cuts on the charged-lepton energy $E_l$ (with or without an
additional cut on the invariant mass $M_X$ of the hadronic system) were used
by the CLEO \cite{Bornheim:2002du}, BABAR \cite{Aubert:2005im,Aubert:2005mg},
and BELLE \cite{Limosani:2005pi} collaborations.
The three other cut scenarios, which were adopted by BABAR
\cite{Aubert:2003zw} and BELLE \cite{Kakuno:2003fk,Bizjak:2005hn} and which
we shall consider here, combine cuts on $E_l$ with cuts on $M_X$, the
invariant mass square $q^2$ of the leptonic system \cite{Bauer:2001rc}, and
the variable $P_+=E_X-|\vec{p}_X|$ \cite{Bosch:2004bt}, where $E_X$ and
$\vec{p}_X$ are the energy and three-momentum of the hadronic system $X_u$,
respectively.
Specifically, they are defined as:
(1) $E_l>1$~GeV, $M_X<1.7$~GeV, and $q^2>8$~GeV$^2$;
(2) $E_l>1$~GeV and $M_X<1.7$~GeV; and
(3) $E_l>1$~GeV and $P_+<0.66$~GeV.
The corresponding fractional decay rates will be denoted as $r_1$, $r_2$, and
$r_3$, respectively.
They all depend on the description of the non-perturbative $b\to B$
transition, for which we shall use the SF and LC approaches as discussed in
the previous section.

Both the SF $F(k_+)$ and the distribution function $f(\xi)$ of the LC approach
depend strongly on the $b$-quark mass and much less on the parameters
$\lambda_1$ and $\lambda_2$, as we shall see below.
For these parameters, we choose $m_b=(4.72\pm0.08)$~GeV,
$\lambda_1=(-0.25\pm0.10)$~GeV$^2$, and $\lambda_2=0.12$~GeV$^2$.
Since the $b$ quark cannot be observed due to confinement, the value of $m_b$
can only be obtained indirectly from measurements other than that of
$B\to X_ul\overline{\nu}_l$. 
The value of $m_b$ depends on the scheme, in which it is defined.
For simplicity, we take $m_b$ to be the pole mass.
The scale-invariant $b$-quark mass in the modified minimal-subtraction
($\overline{\rm MS}$) scheme currently quoted by the
Particle Data Group \cite{Yao:2006px} as
$\overline{m}_b=\overline{m}_b(\overline{m}_b)=(4.20\pm0.07)$~GeV
corresponds to $m_b=(4.78\pm0.08)$~GeV at the two-loop level.
A determination of $m_b$ and $\lambda_1$ by fitting $B\to X_s\gamma$ decay
spectra may be found in Ref.~\cite{Aubert:2005cu}, with the result that
$m_b=\left(4.79_{-0.10}^{+0.06}\right)$~GeV and
$\lambda_1=\left(-0.24_{-0.18}^{+0.09}\right)$~GeV$^2$.
In the analysis of their data \cite{Bizjak:2005hn}, the BELLE Collaboration
used the values $m_b=4.60$~GeV and $\lambda_1=-0.20$~GeV$^2$ within the SF
scheme.
All these values are consistent with our above choice for $m_b$ and
$\lambda_1$.
With these parameters, we calculate the parameters $\overline{\Lambda}$ and
$c$ that fix the SF $F(k_+)$ in Eq.~(\ref{eq:shape}) as well as, via
Eq.~(\ref{eq:musig}), the parameters $a$ and $b$ that fix the distribution
function $f(\xi)$ of the LC approach in Eq.~(\ref{eq:ans}).
In the latter case, we also need as input the parameter $\lambda_2$, which we
fix as described above. 
The values of $\overline{\Lambda}$ and $c$ in Eq.~(\ref{eq:shape}) and those
of $a$ and $b$ in Eq.~(\ref{eq:ans}) are collected in Tables~\ref{tab:one} and
\ref{tab:two}, respectively, for $m_b=4.64$, 4.72, and 4.80~GeV and for
$\lambda_1=-0.35$, $-0.25$, and $-0.15$~GeV$^2$.
\begin{table}[ht]
\begin{center}
\caption{\label{tab:one}%
Values of $\overline{\Lambda}$ (in GeV) and
$c=-3\overline{\Lambda}/\lambda_1-1$ appearing in Eq.~(\ref{eq:shape}) for
various values of $m_b$ (in GeV) and $\lambda_1$ (in GeV$^2$).}
\medskip
\begin{tabular}{|c|ccc|}
\hline
$m_b$ & 4.64 & 4.72 & 4.80 \\
\hline
\backslashbox{$\lambda_1$}{$\overline{\Lambda}$} & 0.6392 & 0.5592 & 0.4792 \\
\hline
$-0.35$ & 2.5021 & 1.6803 & 0.9683 \\
$-0.25$ & 3.9029 & 2.7525 & 1.7556 \\
$-0.15$ & 7.1715 & 5.2541 & 3.5927 \\
\hline
\end{tabular}
\end{center}
\end{table}
\begin{table}[ht]
\begin{center}
\caption{\label{tab:two}%
Values of $a$ and $b$ appearing in Eq.~(\ref{eq:ans}) for various values of
$m_b$ (in GeV) and $\lambda_1$ (in GeV$^2$).}
\medskip
\begin{tabular}{|c|ccc|c|}
\hline
\backslashbox{$\lambda_1$}{$m_b$} & 4.64 & 4.72 & 4.80 & \\
\hline
$-0.35$ & 0.007950 & 0.006940 & 0.005895 & $a$ \\
& 0.8941 & 0.9094 & 0.9245 & $b$ \\
$-0.25$ & 0.005911 & 0.005215 & 0.004493 & $a$ \\
& 0.8861 & 0.9014 & 0.9166 & $b$ \\
$-0.15$ & 0.003604 & 0.003212 & 0.002804 & $a$ \\
& 0.8780 & 0.8934 & 0.9087 & $b$ \\
\hline
\end{tabular}
\end{center}
\end{table}

Before we can present our results for $r_1$, $r_2$, and $r_3$, we need to know
the change of the fully integrated decay rate of $B\to X_ul\overline{\nu}_l$
due to the Fermi motion of the $b$ quark inside the $B$ meson.
Therefore, we write
\begin{equation}
\Gamma(B\to X_ul\overline{\nu}_l)=r_0\Gamma(b\to X_ul\overline{\nu}_l),
\label{eq:r0}
\end{equation}
where $\Gamma(b\to X_ul\overline{\nu}_l)$ is given by Eq.~(\ref{eq:gamtot})
and the deviation of $r_0$ from unity measures the influence of the Fermi
motion.
The results for $r_0$ evaluated in the SF and LC approaches with the fixed
value $\alpha_s=0.22$ are given in Tables~\ref{tab:three} and \ref{tab:four},
respectively, for the same values of $m_b$ and $\lambda_1$ as in
Tables~\ref{tab:one} and \ref{tab:two}.
We see that, in both approaches, $r_0$ is approximately equal to one.
The variation with $m_b$ is very small; $r_0$ mostly depends on $\lambda_1$.
The deviation of $r_0$ from unity is because the factor $m_b$ in
$\Gamma_0$ [see Eq.~(\ref{eq:gam0})] is replaced by $\langle m_b+k_+\rangle^5$
in the SF case and by $\langle\xi m_b\rangle^5$ in the LC case.
It is instructive to approximate these expectation values by their lowest
non-vanishing moments.
In the SF case, we thus obtain for $r_0$:
\begin{equation}
r_0\approx1+\frac{10A_2}{m_b^2},
\end{equation}
where $A_2$ is given in Eq.~(\ref{eq:a2}).
This yields $r_0=1.0374$ for $m_b=4.72$~GeV, almost the same value as in
Table~\ref{tab:three}.
The derivation comes from the higher moments, which must be even smaller.
Of course, these results do not imply that the integrated decay rate is almost
independent of $m_b$.
On the contrary, it is proportional to $m_b^5$ and, therefore, changes with
this factor.
Only the influence of the Fermi motion on this decay rate is small and feebly
depends on $m_b$, as one would expect.
Independently varying $m_b$ and $\lambda_1$, we have
$r_0=1.0353_{-0.0145}^{+0.0153}$.
Table~\ref{tab:four} exhibits a similar pattern for $r_0$ in the LC case.
For our central choice of $m_b$ and $\lambda_1$, it is almost one.
It changes very little with $m_b$ and more with $\lambda_1$.
Over the whole range of $m_b$ and $\lambda_1$, we have
$r_0=1.0044_{-0.0309}^{+0.0297}$.
Approximating $r_0$ by the first two non-vanishing moments, we obtain
\begin{equation}
r_0\approx1+\frac{25}{3}E_b+\frac{20}{3}K_b
=1-\frac{45\lambda_1}{6m_b^2}-\frac{25\lambda_2}{2m_b^2},
\end{equation}
which yields $r_0\approx1.0168$ for our default values of $m_b$, $\lambda_1$,
and $\lambda_2$. 
Comparison with Table~\ref{tab:four} reveals that, in the LC case, the higher
moments are more important than in the SF case.
Since the error of $r_0$ is doubled as compared to the SF case, the error in
the integrated decay rate is also larger.
From Tables~\ref{tab:three} and \ref{tab:four}, we may also conclude that
parton-hadron duality is realized to good approximation for the total decay
rate, $r_0$ being close to unity.
\begin{table}[ht]
\begin{center}
\caption{\label{tab:three}%
Values of $r_0$ appearing in Eq.~(\ref{eq:r0}) evaluated for various values of
$m_b$ (in GeV) and $\lambda_1$ (in GeV$^2$) in the SF approach.}
\medskip
\begin{tabular}{|c|ccc|}
\hline
\backslashbox{$\lambda_1$}{$m_b$} & 4.64 & 4.72 & 4.80 \\
\hline
$-0.35$ & 1.0506 & 1.0484 & 1.0463 \\
$-0.25$ & 1.0369 & 1.0353 & 1.0338 \\
$-0.15$ & 1.0225 & 1.0217 & 1.0208 \\
\hline
\end{tabular}
\end{center}
\end{table}
\begin{table}[ht]
\begin{center}
\caption{\label{tab:four}%
Values of $r_0$ appearing in Eq.~(\ref{eq:r0}) evaluated for various values of
$m_b$ (in GeV) and $\lambda_1$ (in GeV$^2$) in the LC approach.}
\medskip
\begin{tabular}{|c|ccc|}
\hline
\backslashbox{$\lambda_1$}{$m_b$} & 4.64 & 4.72 & 4.80 \\
\hline
$-0.35$ & 1.0350 & 1.0335 & 1.0319 \\
$-0.25$ & 1.0049 & 1.0044 & 1.0041 \\
$-0.15$ & 0.9747 & 0.9755 & 0.9759 \\
\hline
\end{tabular}
\end{center}
\end{table}

Next we present our results for the fractional decay rates $r_1$, $r_2$, and
$r_3$.
For the SF approach, they are listed in Table~\ref{tab:five} for the same
choices of $m_b$ and $\lambda_1$ as above.
The central values are $r_1=0.362$, $r_2=0.676$, and $r_3=0.602$.
The results for the LC approach are given in
Table~\ref{tab:six}, the central values being $r_1=0.360$, $r_2=0.694$, and
$r_3=0.667$.
They are similar to the SF case, expect for $r_3$, which is larger in the LC
case. 
The SF to LC ratios read 1.00, 0.97, and 0.90.
Thus, the fractional decay rates are remarkably similar in the two approaches
and differ only little from the results $r_1=0.34$, $r_2=0.66$, and $r_3=0.57$
obtained in Ref.~\cite{Lange:2005yw}, which were used in
Ref.~\cite{Barberio:2006bi} to determine $|V_{ub}|$ through a global analysis
of the available experimental data.
As expected, the values of $r_1$, $r_2$, and $r_3$ depend much more strongly
on $m_b$ than on $\lambda_1$, both in the SF and LC approaches.
The variations of $r_i$ with these two parameters are larger in the LC
approach than in the SF approach.
If we express these variations as errors, we have
$r_1=0.362_{-0.027}^{+0.024}$, $r_2=0.676_{-0.094}^{+0.064}$, and
$r_3=0.602_{-0.140}^{+0.089}$ in the SF approach and
$r_1=0.360_{-0.029}^{+0.026}$, $r_2=0.694_{-0.200}^{+0.094}$, and
$r_3=0.667_{-0.479}^{+0.098}$ in the LC approach.
We notice that, in the LC approach, $r_3$ becomes abnormally small for
$m_b=4.64$~GeV and $\lambda_1=-0.15$~GeV$^2$.
\begin{table}[ht]
\begin{center}
\caption{\label{tab:five}%
Values of $r_1$, $r_2$, and $r_3$ evaluated for various values of $m_b$ (in
GeV) and $\lambda_1$ (in GeV$^2$) in the SF approach.}
\medskip
\begin{tabular}{|c|ccc|c|}
\hline
\backslashbox{$\lambda_1$}{$m_b$} & 4.64 & 4.72 & 4.80 & \\
\hline
& 0.3438 & 0.3659 & 0.3860 & $r_1$ \\
$-0.35$ & 0.6283 & 0.6888 & 0.7398 & $r_2$ \\
& 0.5411 & 0.6223 & 0.6908 & $r_3$ \\
& 0.3386 & 0.3617 & 0.3824 & $r_1$ \\
$-0.25$ & 0.6082 & 0.6763 & 0.7330 & $r_2$ \\
& 0.5076 & 0.6016 & 0.6796 & $r_3$ \\
& 0.3347 & 0.3586 & 0.3795 & $r_1$ \\
$-0.15$ & 0.5828 & 0.6633 & 0.7290 & $r_2$ \\
& 0.4614 & 0.5779 & 0.6724 & $r_3$ \\
\hline
\end{tabular}
\end{center}
\end{table}
\begin{table}[ht]
\begin{center}
\caption{\label{tab:six}%
Values of $r_1$, $r_2$, and $r_3$ evaluated for various values of $m_b$ (in
GeV) and $\lambda_1$ (in GeV$^2$) in the LC approach.}
\medskip
\begin{tabular}{|c|ccc|c|}
\hline
\backslashbox{$\lambda_1$}{$m_b$} & 4.64 & 4.72 & 4.80 & \\
\hline
& 0.3476 & 0.3674 & 0.3856 & $r_1$ \\
$-0.35$ & 0.6201 & 0.7305 & 0.7878 & $r_2$ \\
& 0.5668 & 0.6997 & 0.7657 & $r_3$ \\
& 0.3395 & 0.3601 & 0.3788 & $r_1$ \\
$-0.25$ & 0.5578 & 0.6942 & 0.7743 & $r_2$ \\
& 0.4584 & 0.6674 & 0.7520 & $r_3$ \\
& 0.3313 & 0.3527 & 0.3720 & $r_1$ \\
$-0.15$ & 0.4945 & 0.6300 & 0.7557 & $r_2$ \\
& 0.1886 & 0.6192 & 0.7357 & $r_3$ \\
\hline
\end{tabular}
\end{center}
\end{table}

\begin{figure}[ht]
\begin{center}
\includegraphics[bb=109  104  530  672,width=0.9\textwidth]{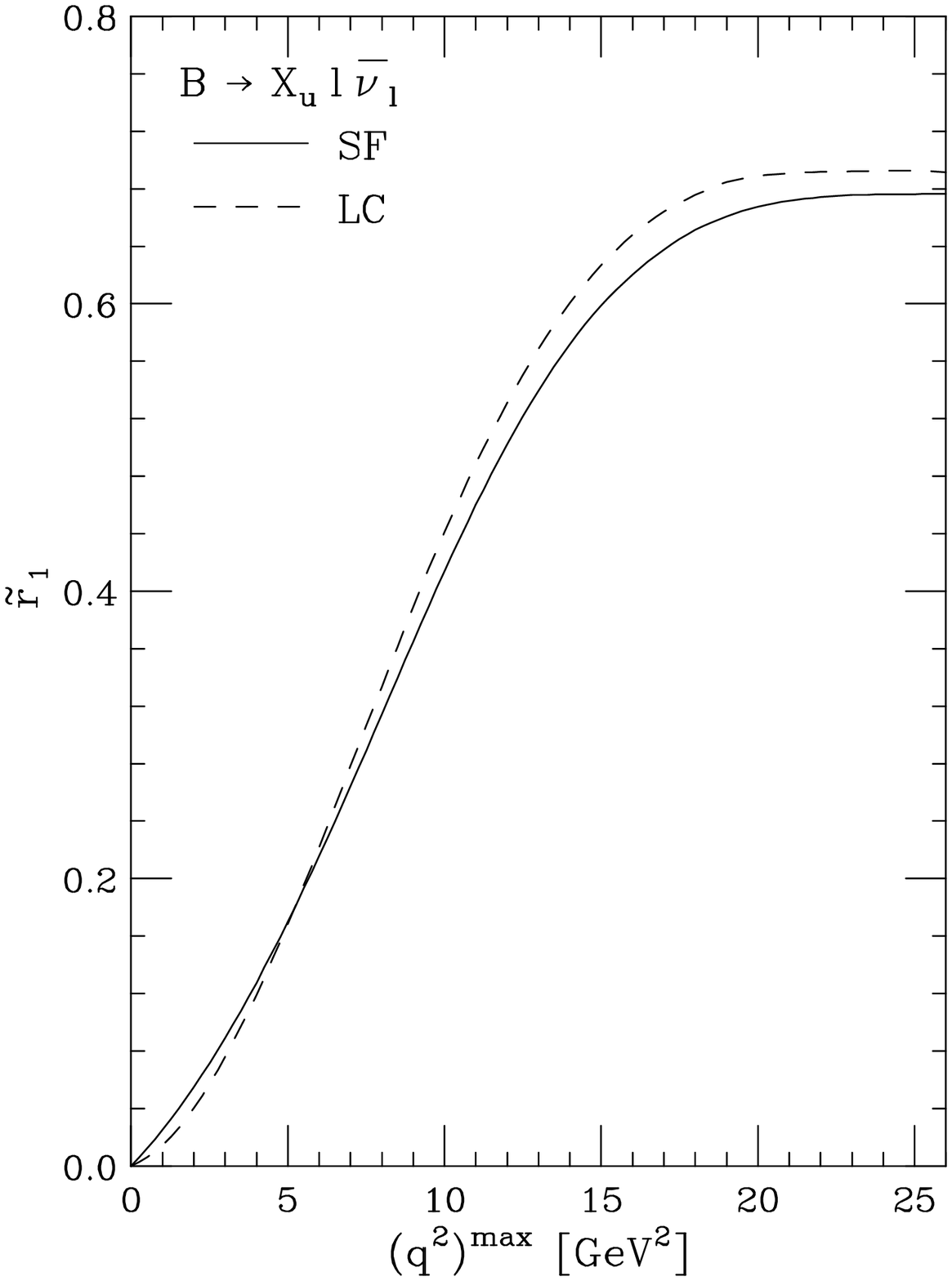}
\caption{\label{fig:one}%
Fractional decay rate $\tilde r_1$ defined in Eq.~(\ref{eq:r1}) evaluated as
a function of $(q^2)^{\rm max}$ in the SF (solid line) and LC (dashed line)
approaches.}
\end{center}
\end{figure}

\begin{figure}[ht]
\begin{center}
\includegraphics[bb=109  104  530  672,width=0.9\textwidth]{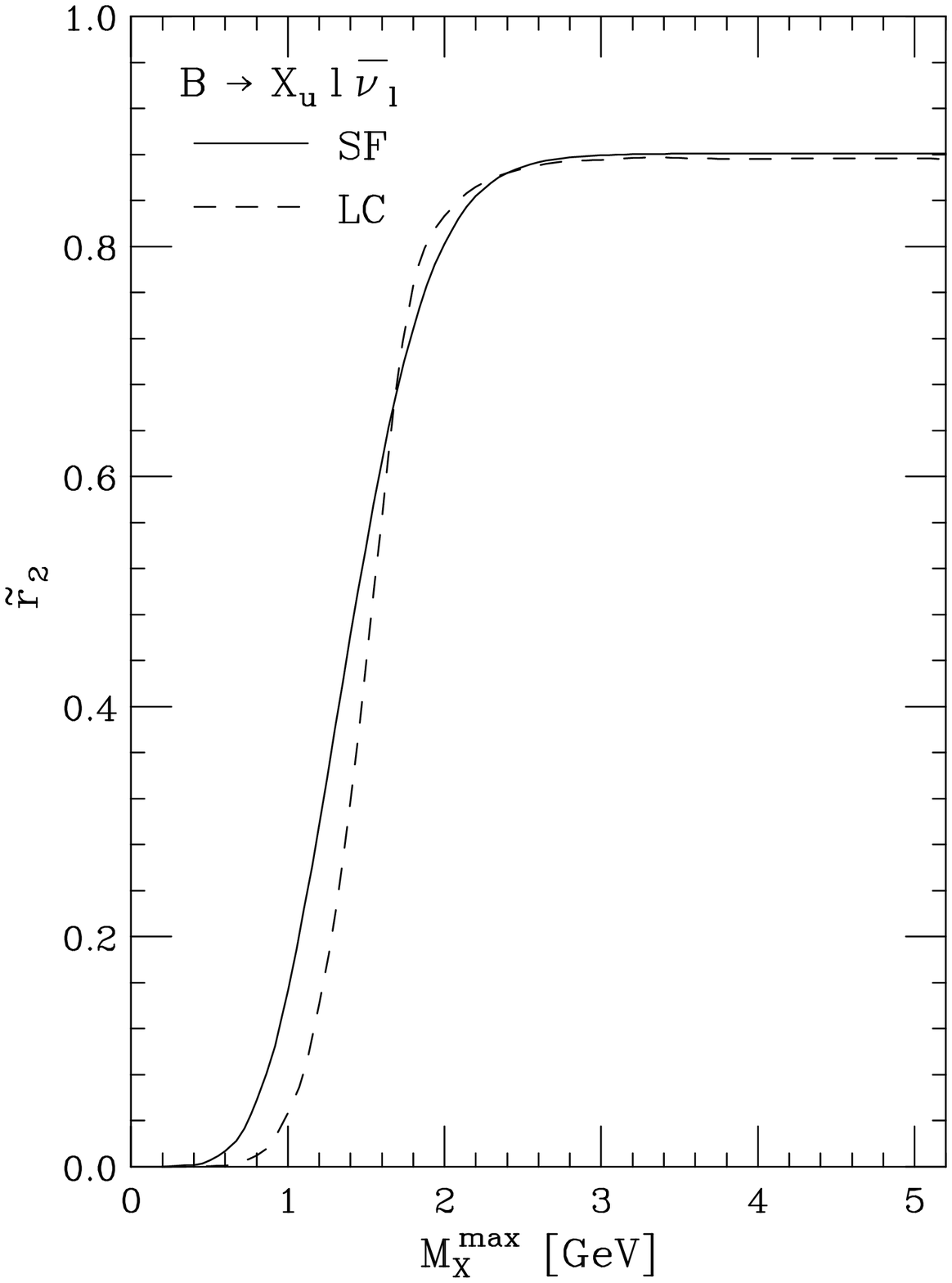}
\caption{\label{fig:two}%
Fractional decay rate $\tilde r_2$ defined in Eq.~(\ref{eq:r2}) evaluated as
a function of $M_X^{\rm max}$ in the SF (solid line) and LC (dashed line)
approaches.}
\end{center}
\end{figure}

\begin{figure}[ht]
\begin{center}
\includegraphics[bb=109  104  541  672,width=0.9\textwidth]{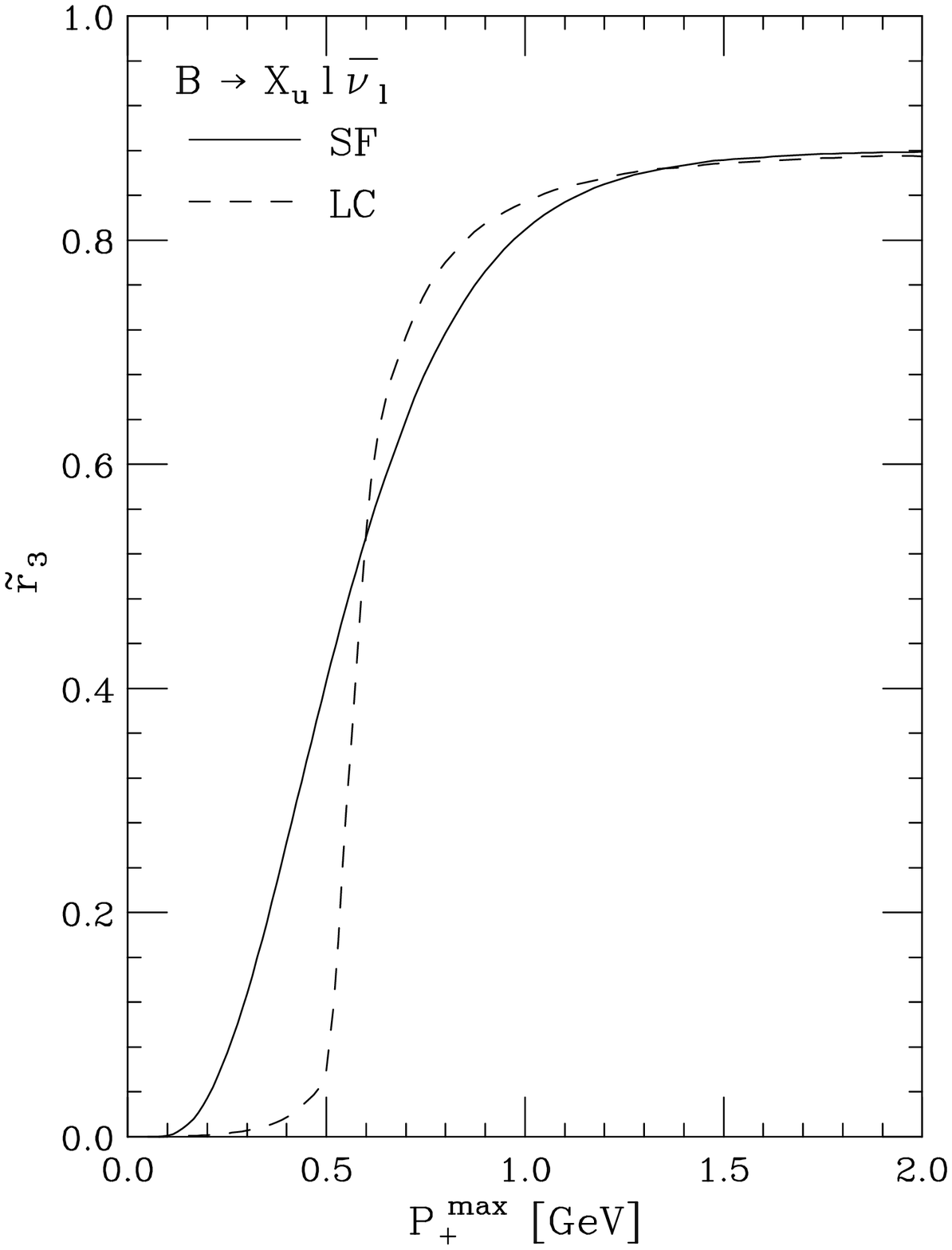}
\caption{\label{fig:three}%
Fractional decay rate $\tilde r_3$ defined in Eq.~(\ref{eq:r3}) evaluated as
a function of $P_+^{\rm max}$ in the SF (solid line) and LC (dashed line)
approaches.}
\end{center}
\end{figure}

The similarity of the fractional decay rates $r_1$, $r_2$, and $r_3$ in the
two approaches considered here might be related to a fortunate choice of the
cut parameters $E_l^{\rm min}$, $M_X^{\rm max}$, $(q^2)^{\rm min}$, and
$P_+^{\rm max}$, whereas the distributions in $M_X$, $q^2$, and $P_+$ for a
fixed value of $E_l^{\rm min}=1$~GeV say, could differ significantly.
To elucidate this point, we calculate the partial decay fractions $r_1$,
$r_2$, and $r_3$ as functions of the cut parameters $M_X^{\rm max}$,
$(q^2)^{\rm min}$, and $P_+^{\rm max}$ for the default values of the input
parameters, $m_b=4.72$~GeV ($\overline{\Lambda}=0.5592$~GeV) and
$\lambda_1=-0.25$~GeV$^2$.
For this purpose, we define
\begin{eqnarray}
\tilde r_1\left((q^2)^{\rm max}\right)
&=&\frac{1}{\Gamma}\int_0^{(q^2)^{\rm max}}dq^2
\left.\frac{d\Gamma}{dq^2}\right|_{E_l>1~{\rm GeV},\ M_X<1.7~{\rm GeV}},
\label{eq:r1}\\
\tilde r_2\left(M_X^{\rm max}\right)
&=&\frac{1}{\Gamma}\int_0^{M_X^{\rm max}}dM_X
\left.\frac{d\Gamma}{dM_X}\right|_{E_l>1~{\rm GeV}},
\label{eq:r2}\\
\tilde r_3\left(P_+^{\rm max}\right)
&=&\frac{1}{\Gamma}\int_0^{P_+^{\rm max}}dP_+
\left.\frac{d\Gamma}{dP_+}\right|_{E_l>1~{\rm GeV}},
\label{eq:r3}
\end{eqnarray}
which are related to $r_1$, $r_2$, and $r_3$ as
\begin{eqnarray}
r_1&=&\tilde r_1\left(26~{\rm GeV}^2\right)
-\tilde r_1\left(8~{\rm GeV}^2\right),
\nonumber\\
r_2&=&\tilde r_2(1.7~{\rm GeV}),
\nonumber\\
r_3&=&\tilde r_3(0.66~{\rm GeV}).
\end{eqnarray}
In Fig.~\ref{fig:one}, $\tilde r_1$ is plotted as a function of
$(q^2)^{\rm max}$ for the SF (solid line) and LC (dashed line) approaches.
We observe that the difference between the two approaches is rather small over
the whole range of $(q^2)^{\rm max}$, way up to 25~GeV$^2$.
Later, when we compare with experimental measurements, we shall see that the
same holds true for the normalized distribution $(1/\Gamma)d\Gamma/dq^2$ with
the above cuts on $E_l$ and $M_X$.
The situation is very similar for $\tilde r_2$, which is shown as a function
of $M_X^{\rm max}$ in Fig.~\ref{fig:two}.
Here, the difference between the two approaches is appreciable only for small
values of $M_X^{\rm max}$, for $M_X^{\rm max}\alt1.5$~GeV.
The situation is very different for $\tilde r_3$, which is depicted as a
function of $P_+^{\rm max}$ for the two approaches in Fig.~\ref{fig:three}.
We observe that the two distributions coincide at
$P_+^{\rm max}\approx0.6$~GeV, where their slopes are very different, however.
The result of the LC approach is somewhat larger above this value of
$P_+^{\rm max}$, way up to $P_+^{\rm max}\approx1.2$~GeV, while is
significantly smaller below.
As we shall illustrate below, this may be understood by considering the
normalized $P_+$ distribution $(1/\Gamma)d\Gamma/dP_+$ with the cut
$E_l>1$~GeV, which is very different for the two approaches.
It turns out that the choice of $E_l^{\rm min}$ is not responsible for this
difference.

\begin{figure}[ht]
\begin{center}
\includegraphics[bb=105  104  541  668,width=0.9\textwidth]{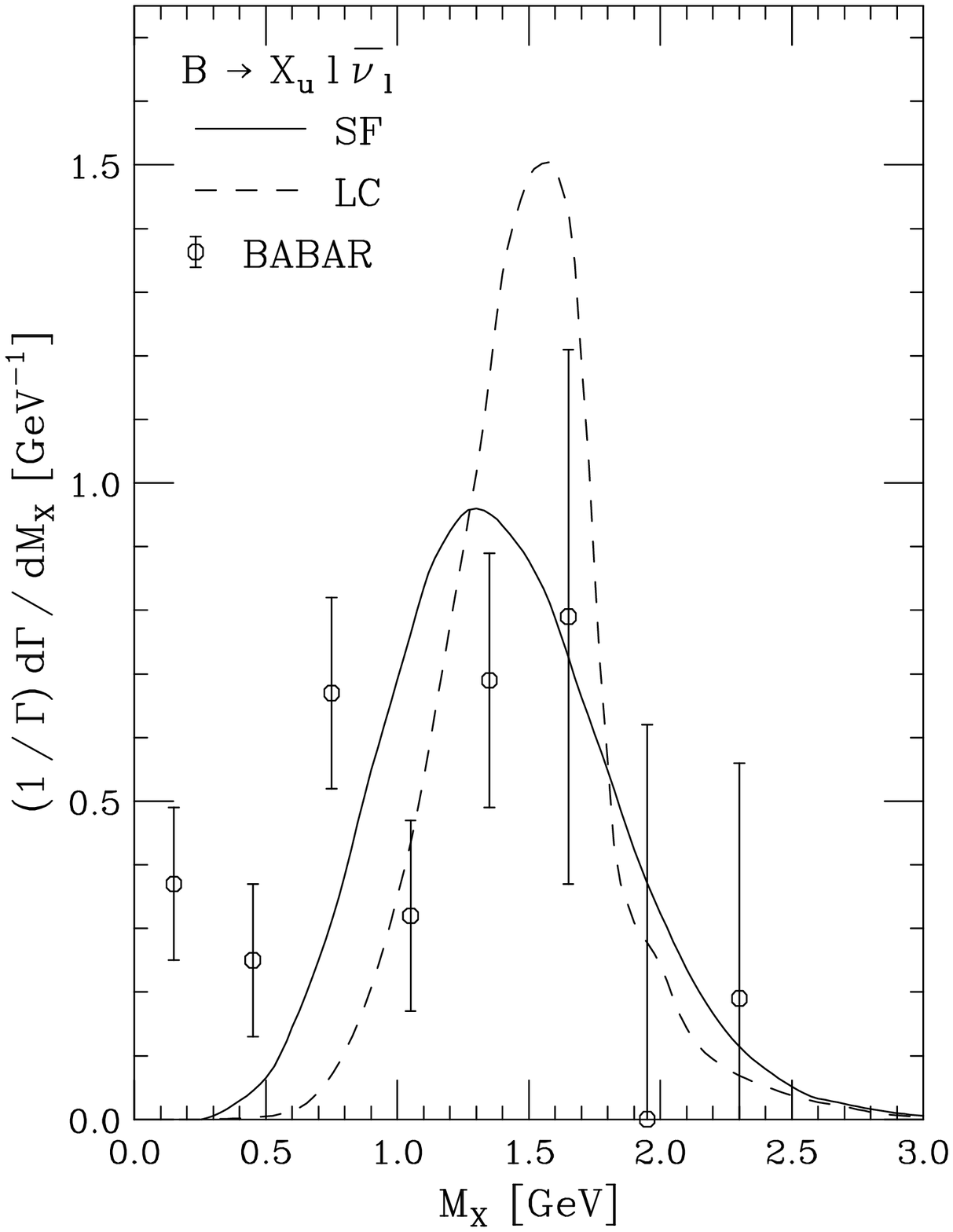}
\caption{\label{fig:four}%
The decay distribution $(1/\Gamma)d\Gamma/dM_X$ with $E_l>1$~GeV normalized to
unity in the signal region ($M_X<2.5$~GeV) as predicted in the SF (solid line)
and LC (dashed line) approaches is compared with BABAR data
\cite{Aubert:2006qi}.}
\end{center}
\end{figure}

\begin{figure}[ht]
\begin{center}
\includegraphics[bb=105  104  541  672,width=0.9\textwidth]{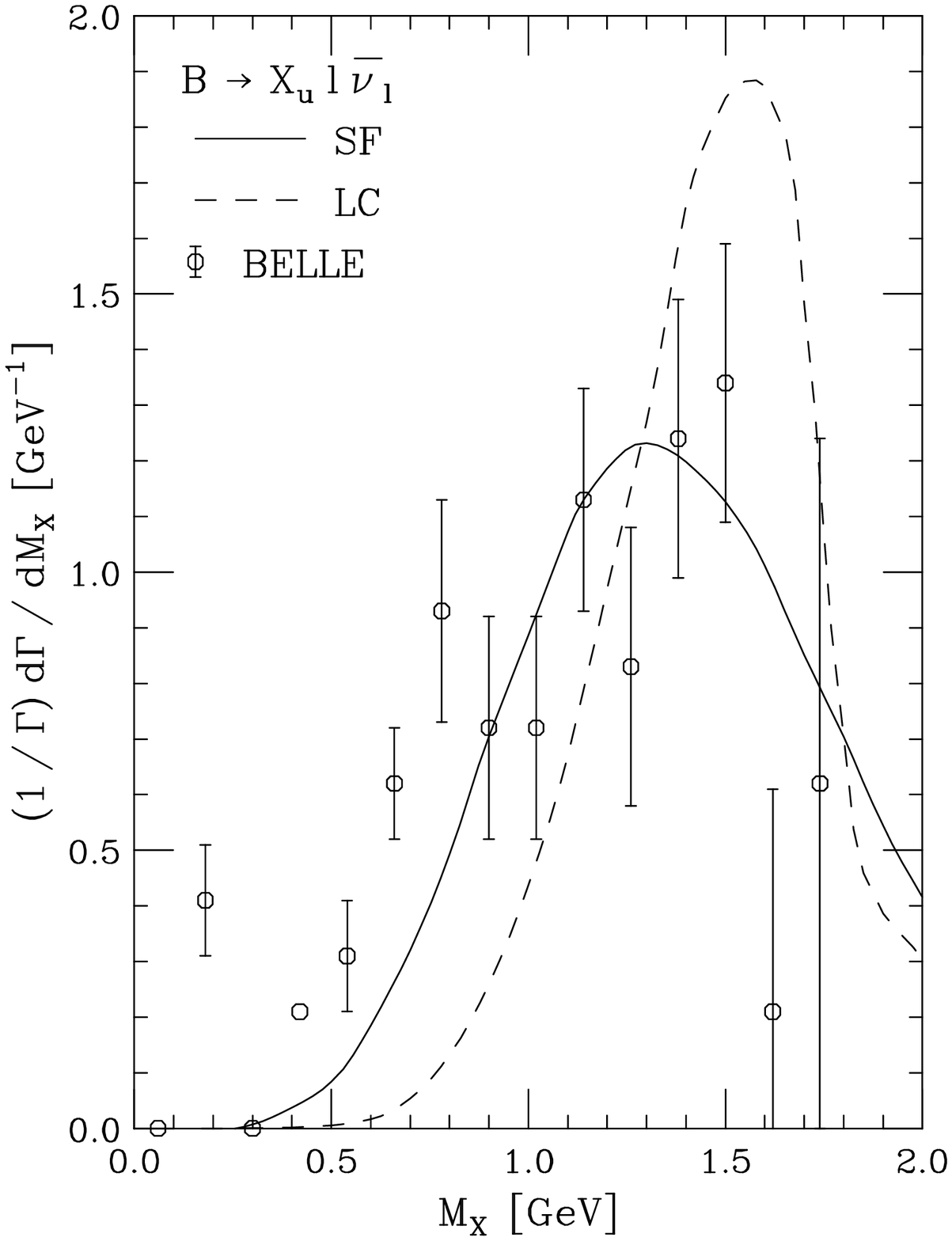}
\caption{\label{fig:five}%
The decay distribution $(1/\Gamma)d\Gamma/dM_X$ with $E_l>1$~GeV normalized to
unity in the signal region ($M_X<1.7$~GeV) as predicted in the SF (solid line)
and LC (dashed line) approaches is compared with BELLE data
\cite{Bizjak:2005hn}.}
\end{center}
\end{figure}

\begin{figure}[ht]
\begin{center}
\includegraphics[bb=105  104  546  672,width=0.9\textwidth]{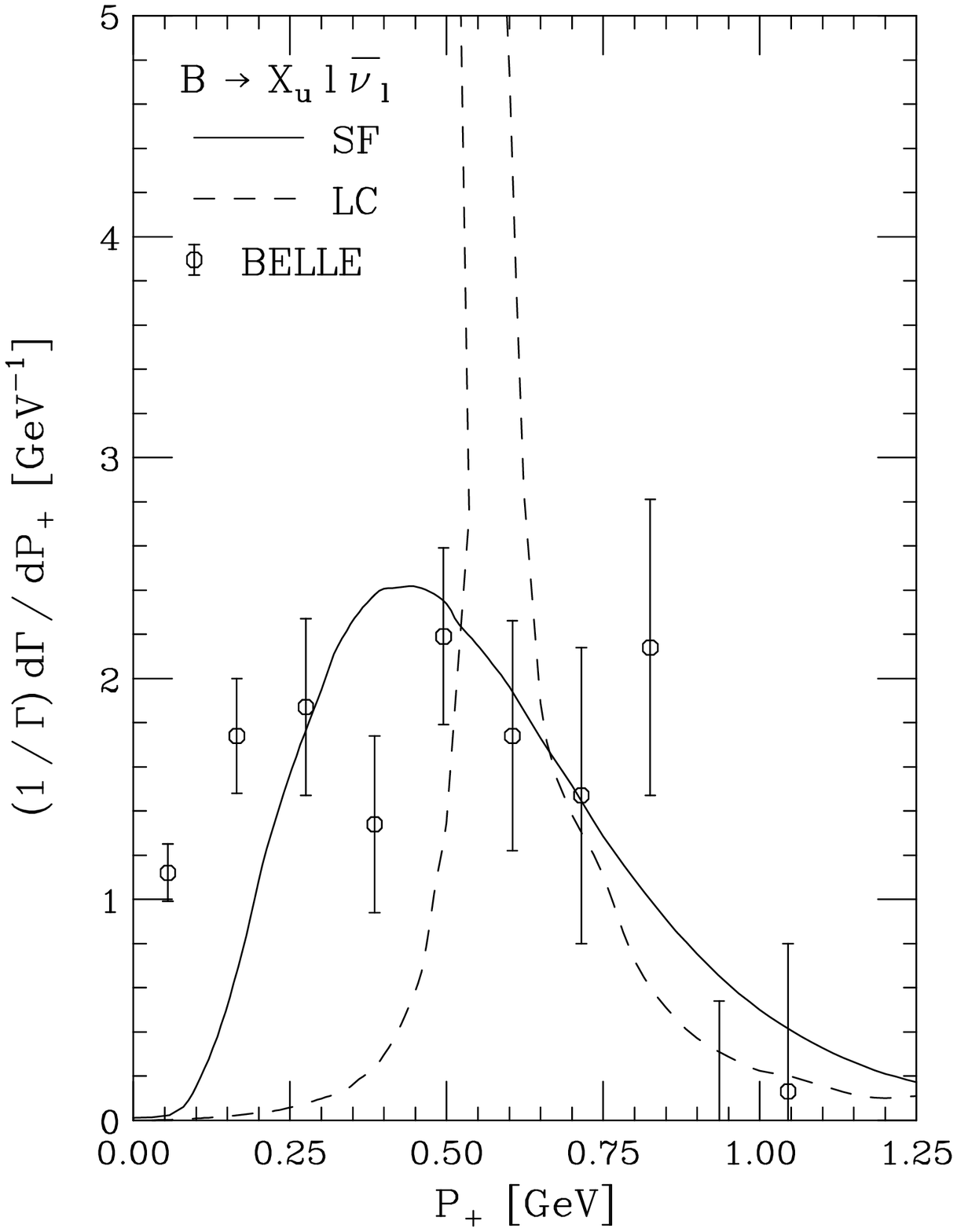}
\caption{\label{fig:six}%
The decay distribution $(1/\Gamma)d\Gamma/dP_+$ with $E_l>1$~GeV normalized to
unity in the signal region ($P_+<0.66$~GeV) as predicted in the SF (solid
line) and LC (dashed line) approaches is compared with BELLE data
\cite{Bizjak:2005hn}.}
\end{center}
\end{figure}

\begin{figure}[ht]
\begin{center}
\includegraphics[bb=78  104  539  672,width=0.9\textwidth]{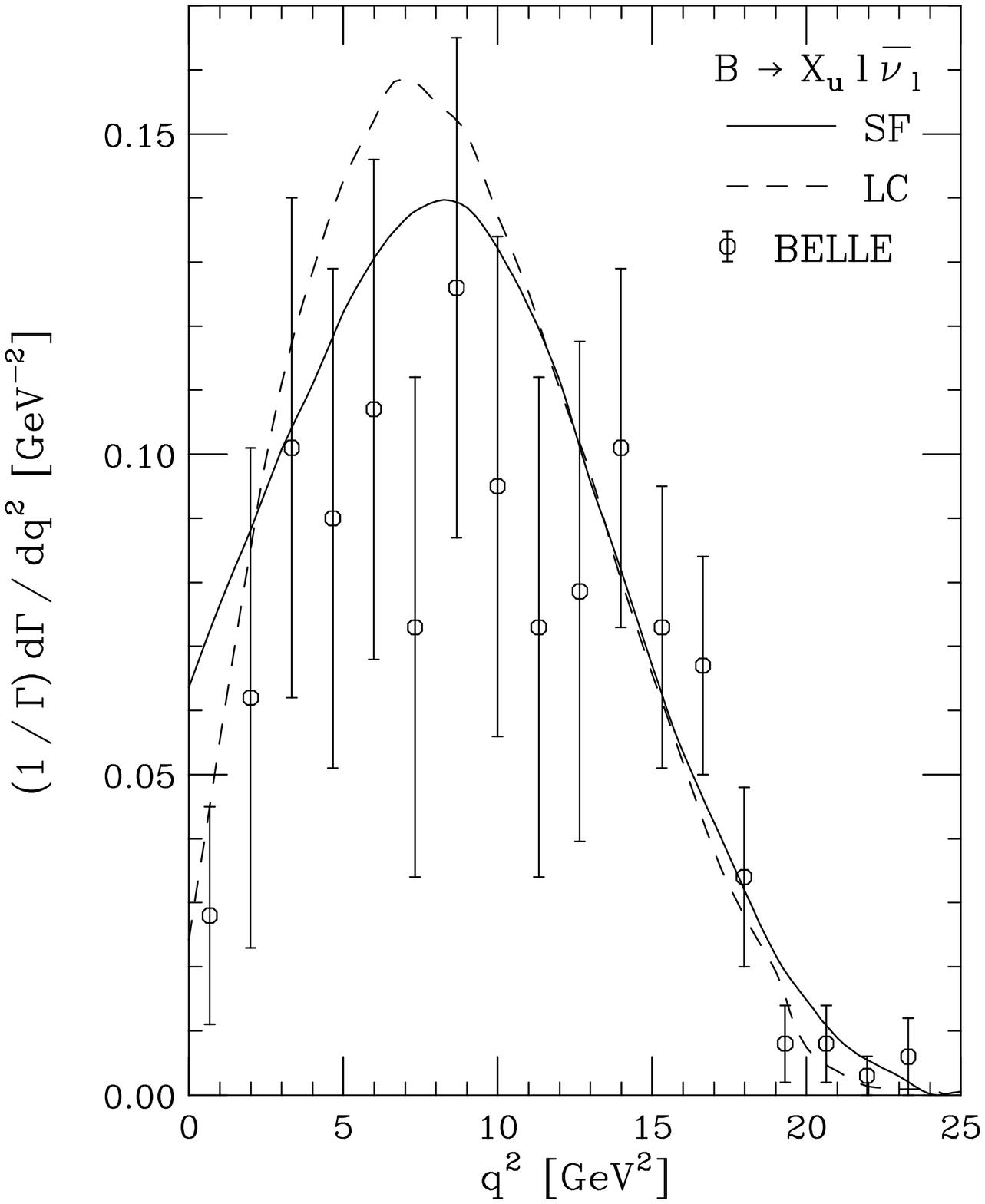}
\caption{\label{fig:seven}%
The decay distribution $(1/\Gamma)d\Gamma/dq^2$ with $E_l>1$~GeV normalized to
unity in the signal region ($M_X<1.7$~GeV and $q^2>8$~GeV$^2$) as predicted in
the SF (solid line) and LC (dashed line) approaches is compared with BELLE
data \cite{Bizjak:2005hn}.}
\end{center}
\end{figure}

As for measurements of fractional decay rates $R(\Delta\Phi)$, experimental
data for the normalized distributions $(1/\Gamma)d\Gamma/dM_X$ and
$(1/\Gamma)d\Gamma/dP_+$ with cuts on $E_l$ have been published and can be
compared to the respective distributions evaluated in the SF and LC approaches
(see Ref.~\cite{Aglietti:2006yb} for a similar comparison).
Specifically, $(1/\Gamma)d\Gamma/dM_X$ distributions with $E_l>1$~GeV have
been published by BABAR \cite{Aubert:2003zw,Aubert:2006qi} and BELLE
\cite{Bizjak:2005hn}.
In Figs.~\ref{fig:four} and \ref{fig:five}, we compare these measured
distributions to our predictions in the SF and LC approaches.
Both the measured and predicted distributions are normalized to unity in the
signal region, which is defined by $M_X<2.5$~GeV for BABAR
\cite{Aubert:2006qi} and by $M_X<1.7$~GeV for BELLE \cite{Bizjak:2005hn}.
From Figs.~\ref{fig:four} and \ref{fig:five}, we see that the predictions in
the SF approach are in reasonable agreement with both measurements, whereas
the distributions of the LC approach are much too narrow and their peaks are
much higher than in the measured distributions.
A similar comparison is performed in Fig.~\ref{fig:six} for the normalized
distribution $(1/\Gamma)d\Gamma/dP_+$ with $E_l>1$~GeV measured by BELLE
\cite{Bizjak:2005hn}. 
Both the measured and calculated distributions are normalized to unity in the
signal region defined by $P_+<0.66$~GeV.
Again, the distribution in the SF approach agrees more or less with the
experimental data, whereas the one in the LC approach is much too narrow.
BELLE \cite{Bizjak:2005hn} also presented experimental data on the
normalized distribution $(1/\Gamma)d\Gamma/dq^2$ with $E_l>1$~GeV normalized
to unity in the signal region defined by $M_X<1.7$~GeV and $q^2>8$~GeV$^2$.
These are compared in Fig.~\ref{fig:seven} with the predictions based on the
SF and LC approaches.
Here, the two theoretical distributions are very similar and both agree with
the measurement reasonably well.

\begin{figure}[ht]
\begin{center}
\includegraphics[bb=109  104  530  672,width=0.9\textwidth]{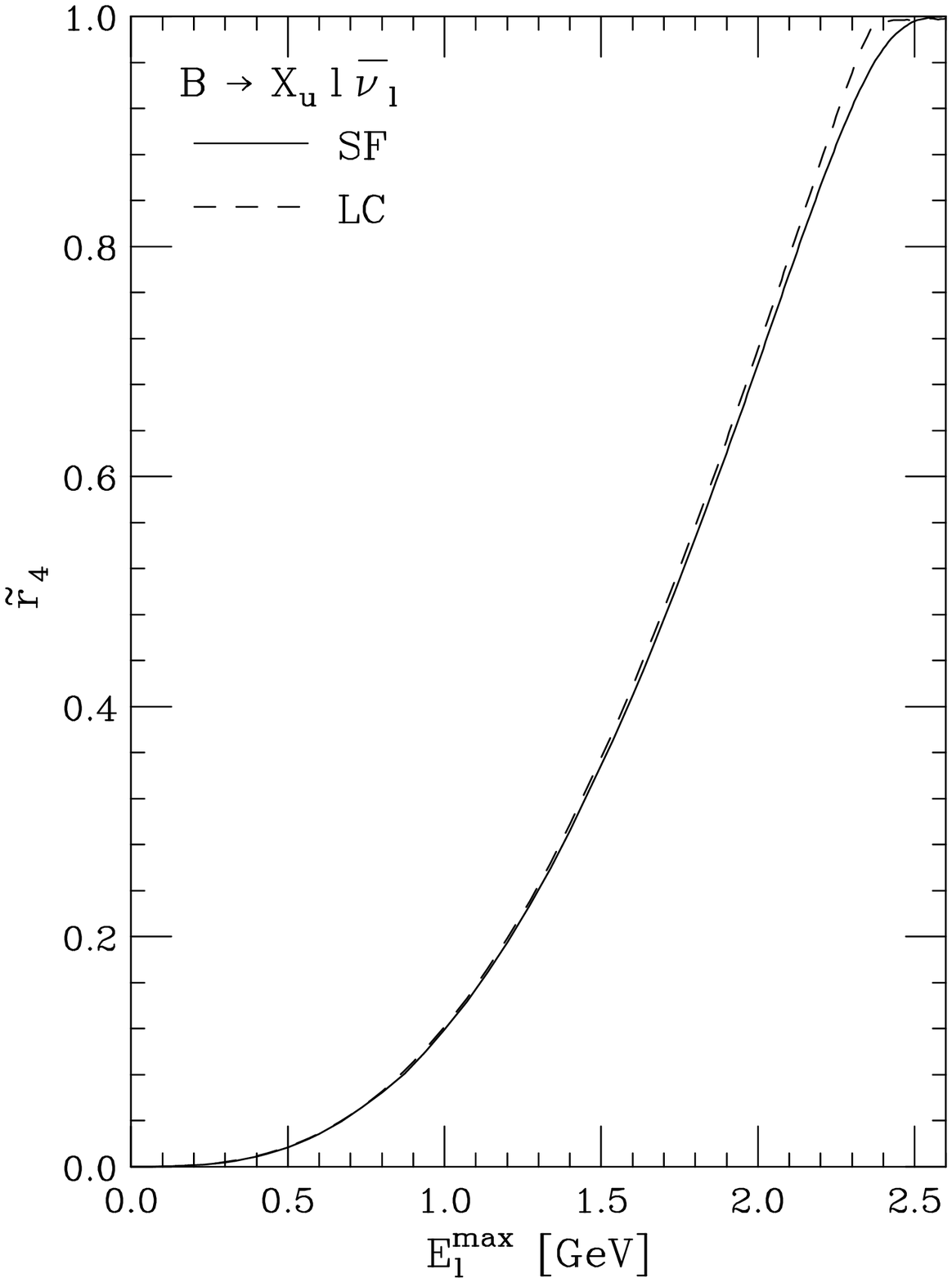}
\caption{\label{fig:eight}%
Fractional decay rate $\tilde r_4$ defined in Eq.~(\ref{eq:r4}) evaluated as
a function of $E_l^{\rm max}$ in the SF (solid line) and LC (dashed line)
approaches.}
\end{center}
\end{figure}

\begin{figure}[ht]
\begin{center}
\includegraphics[bb=78  104  530  672,width=0.9\textwidth]{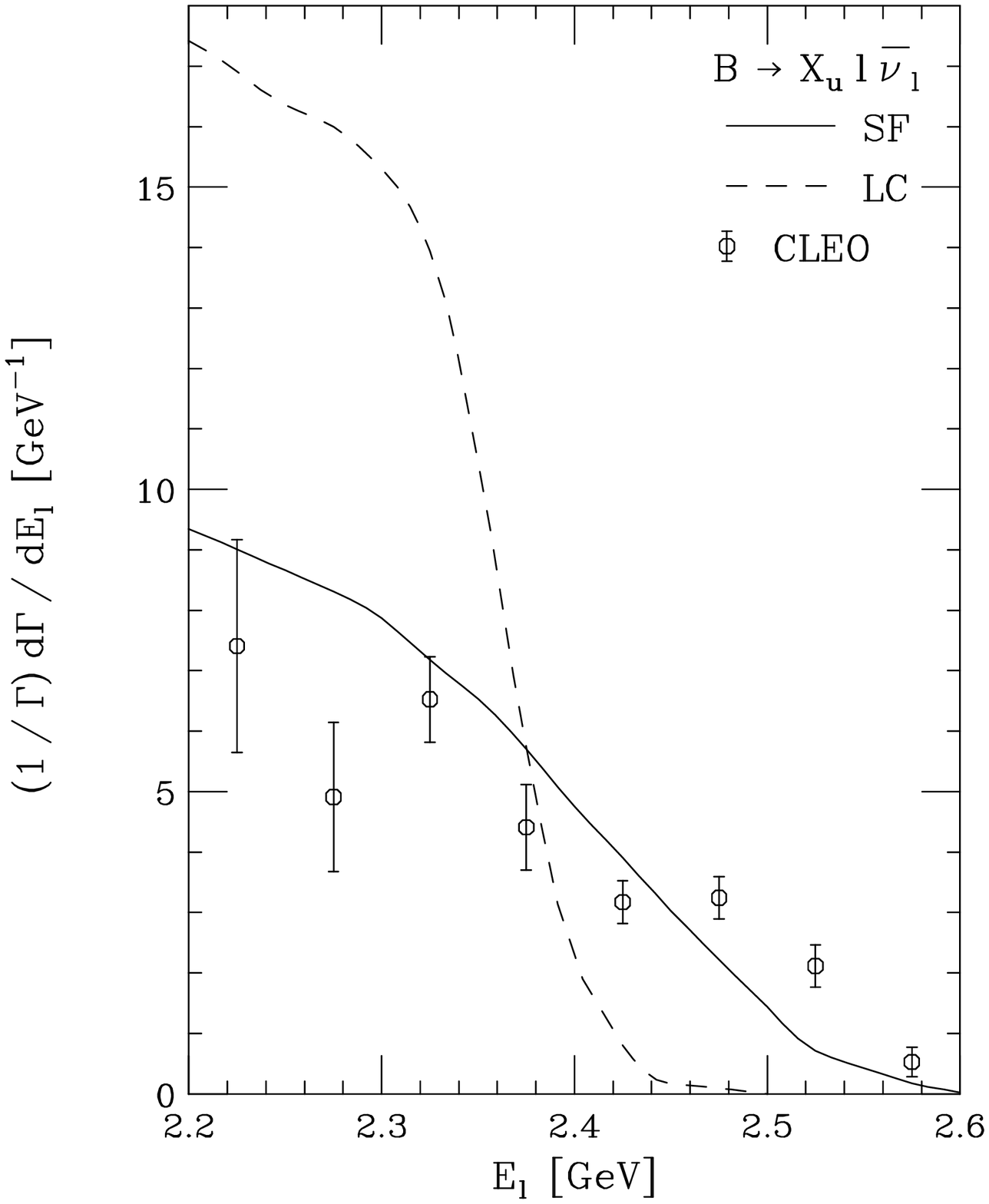}
\caption{\label{fig:nine}%
The decay distribution $(1/\Gamma)d\Gamma/dE_l$ normalized to unity in the
signal region (2.30~GeV${}<E_l<2.60$~GeV) as predicted in the SF (solid line)
and LC (dashed line) approaches is compared with CLEO data
\cite{Bornheim:2002du}.}
\end{center}
\end{figure}

\begin{figure}[ht]
\begin{center}
\includegraphics[bb=78  104  541  672,width=0.9\textwidth]{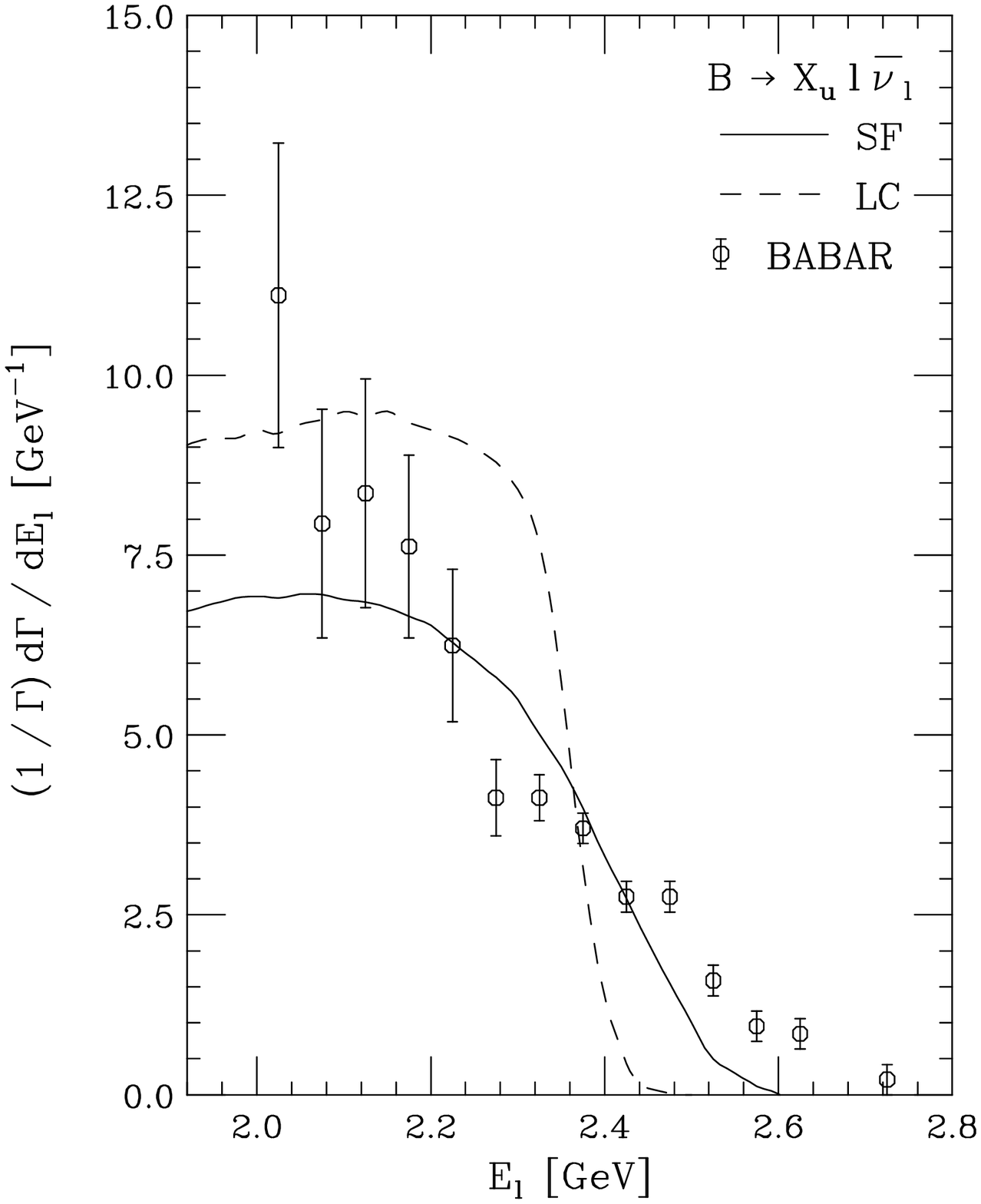}
\caption{\label{fig:ten}%
The decay distribution $(1/\Gamma)d\Gamma/dE_l$ normalized to unity in the
signal region (2.25~GeV${}<E_l<2.60$~GeV) as predicted in the SF (solid line)
and LC (dashed line) approaches is compared with BABAR data
\cite{Aubert:2005mg}.}
\end{center}
\end{figure}

\begin{figure}[ht]
\begin{center}
\includegraphics[bb=78  104  541  672,width=0.9\textwidth]{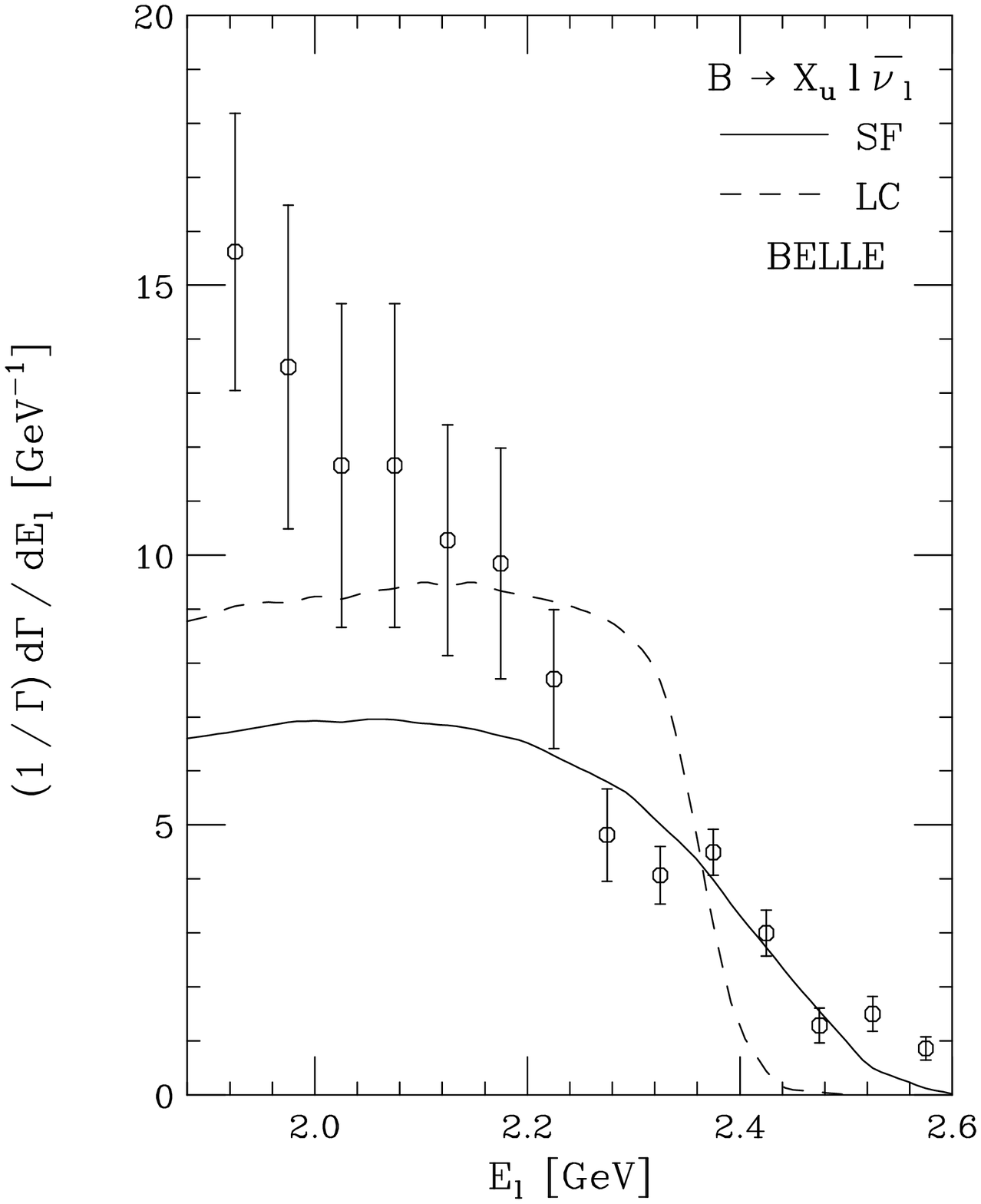}
\caption{\label{fig:eleven}%
The decay distribution $(1/\Gamma)d\Gamma/dE_l$ normalized to unity in the
signal region (2.25~GeV${}<E_l<2.60$~GeV) as predicted in the SF (solid line)
and LC (dashed line) approaches is compared with BELLE data
\cite{Limosani:2005pi}.}
\end{center}
\end{figure}

Finally, we turn to the charged-lepton energy distribution $d\Gamma/dE_l$.
In analogy to Eqs.~(\ref{eq:r1})--(\ref{eq:r3}), we define the fractional
decay rate
\begin{equation}
\tilde r_4\left(E_l^{\rm max}\right)=\frac{1}{\Gamma}
\int_0^{E_l^{\rm max}}dE_l\frac{d\Gamma}{dE_l}.
\label{eq:r4}
\end{equation}
In Tables~\ref{tab:seven} and \ref{tab:eight}, we present the values of
$\tilde r_4(2.3~{\rm GeV})$ evaluated for various values of $m_b$ and
$\lambda_1$ in the SF and LC approaches, respectively.
We notice that, for given values of $m_b$ and $\lambda_1$, the results in the
two approaches differ appreciably.
In particular, $\tilde r_4(2.3~{\rm GeV})$ depends much more strongly on $m_b$
for fixed $\lambda_1$ and vice versa in the LC approach as compared to the SF
approach.
This is due to the fact that the $E_l$ distribution falls off much more
rapidly towards the threshold at $E_l^{\rm max}=M_B/2$ in the LC approach as
compared to the SF approach (see Figs.~\ref{fig:nine}--\ref{fig:eleven}).
Of course, this effect diminishes if $E_l^{\rm max}$ is taken to be smaller
than 2.3~GeV.
In this case, also the sensitivity of $\tilde r_4\left(E_l^{\rm max}\right)$
on $m_b$ and $\lambda_1$ is reduced.
The SF result for $\tilde r_4(2.3~{\rm GeV})$ agrees quite well with the value
used by CLEO \cite{Bornheim:2002du} to determine $|V_{ub}|$ from the data
points in the range 2.3~GeV${}<E_l<{}$2.6~GeV.
\begin{table}[ht]
\begin{center}
\caption{\label{tab:seven}%
Values of $\tilde r_4(2.3~{\rm GeV})$ defined in Eq.~(\ref{eq:r4}) evaluated
for various values of $m_b$ (in GeV) and $\lambda_1$ (in GeV$^2$) in the SF
approach.}
\medskip
\begin{tabular}{|c|ccc|}
\hline
\backslashbox{$\lambda_1$}{$m_b$} & 4.64 & 4.72 & 4.80 \\
\hline
$-0.35$ & 0.9307 & 0.9099 & 0.8874 \\
$-0.25$ & 0.9419 & 0.9210 & 0.8978 \\
$-0.15$ & 0.9558 & 0.9347 & 0.9107 \\
\hline
\end{tabular}
\end{center}
\end{table}
\begin{table}[ht]
\begin{center}
\caption{\label{tab:eight}%
Values of $\tilde r_4(2.3~{\rm GeV})$ defined in Eq.~(\ref{eq:r4}) evaluated
for various values of $m_b$ (in GeV) and $\lambda_1$ (in GeV$^2$) in the LC
approach.}
\medskip
\begin{tabular}{|c|ccc|}
\hline
\backslashbox{$\lambda_1$}{$m_b$} & 4.64 & 4.72 & 4.80 \\
\hline
$-0.35$ & 0.9646 & 0.9385 & 0.9106 \\
$-0.25$ & 0.9780 & 0.9524 & 0.9241 \\
$-0.15$ & 0.9910 & 0.9669 & 0.9382 \\
\hline
\end{tabular}
\end{center}
\end{table}
In Fig.~\ref{fig:eight}, $\tilde r_4$ is displayed as a function of 
$E_l^{\rm max}$ for the SF (solid line) and LC (dashed line) approaches.
We observe that, as $E_l^{\rm max}$ approaches its kinematical upper limit,
the LC result is saturated appreciably earlier than the SF one.
This would lead to an according difference in the value of $|V_{ub}|$
extracted from the data if a large $E_l^{\rm max}$ cut were imposed.
For $E_l^{\rm max}\alt2$~GeV, the SF and LC results for $\tilde r_4$ are very
similar.

In Figs.~\ref{fig:nine}, \ref{fig:ten}, and \ref{fig:eleven}, we compare the
normalized $E_l$ distributions $(1/\Gamma)d\Gamma/dE_l$ predicted by the
SF and LC approaches with measurements by CLEO \cite{Bornheim:2002du}, BABAR
\cite{Aubert:2005mg}, and BELLE \cite{Limosani:2005pi}, respectively.
Both the measured and calculated distributions are normalized to unity in the
signal region, which is defined by $E_l>2.30$~GeV for CLEO
\cite{Bornheim:2002du} and by $E_l>2.25$~GeV for BABAR \cite{Aubert:2005mg}
and BELLE \cite{Limosani:2005pi}.
In the signal region, where the background from $b\to c$ transitions is
expected to be minimal, the SF results agree with the CLEO, BABAR, and BELLE
data quite satisfactorily, while the LC results are clearly disfavored.
In fact, the $E_l$ distributions of the LC approach drop off much too strongly
towards the threshold at $E_l=M_B/2$ and deviate from the data throughout the
signal region.
This disagreement again points to the inadequacy of the LC approach to
describe the non-perturbative effects in $B\to X_ul\overline{\nu}_l$ decays,
which was already noticed for the $M_X$ and $P_+$ distributions in
Figs.~\ref{fig:four}--\ref{fig:six}.
Finally, we should note that, in Figs.~\ref{fig:nine}--\ref{fig:eleven}, the
theoretical predictions refer to the rest frame of the $B$ meson, while the
experimental data refer to that of the $\Upsilon(4S)$ meson.
However, since the motion of the $B$ mesons in the $\Upsilon(4S)$ rest frame
is non-relativistic, this mismatch is rather insignificant in comparison with
the experimental errors.

\section{\label{sec:four}Conclusions}

We studied non-perturbative effects on $B\to X_ul\overline{\nu}_l$ decays due
to the motion of the $b$ quark inside the $B$ meson adopting two approaches
frequently discussed in the literature, namely the shape-function formalism
and the parton model in the light-cone limit.
While these effects are generally small for the total decay rate, they may
become substantial once kinematic acceptance cuts are applied.
In fact, such acceptance cuts are indispensable in practice in order to
suppress the overwhelming background from $B\to X_cl\overline{\nu}_l$ decays.
We considered three cut scenarios, involving the invariant mass $M_X$ of the
hadronic system $X_u$, the variable $P_+=E_X-|\vec{p}_X|$ related to the
energy $E_X$ and the three-momentum $\vec{p}_X$ of $X_u$, the invariant mass
square $q^2$ of the leptonic system, and the charged-lepton energy $E_l$,
that were adopted in recent experimental analyses by the CLEO, BABAR, and
BELLE collaborations.
Comparisons with decay distributions in $M_X$, $P_+$, and $E_l$ measured in
these experiments disfavor the light-cone approach.

\section*{Acknowledgments}

The work of B.A.K. and G.K. was supported in part by the German Federal
Ministry of Education and Research (BMBF) through Grant No.\ 05~HT6GUA.
The work of J.-F.Y. was supported in part by the National Natural Science
Foundation of China through Grant Nos.\ 10205004 and 10475028.

\end{document}